\documentclass[12pt,a4paper,final]{article}

\usepackage{amsmath}
\usepackage{amssymb}
\usepackage{amsthm}
\usepackage{exscale}
\usepackage[mathscr]{eucal}
\usepackage[normalem]{ulem}
\usepackage{floatrow}
\usepackage{bm}
\usepackage{graphicx}
\usepackage{color}
\usepackage{dsfont} 
\usepackage{lineno}
\usepackage[font={small}]{caption}
\usepackage{parskip}

\usepackage[numbers,sort&compress]{natbib}
\usepackage[utf8]{inputenc} 

\usepackage{geometry}
\usepackage{xcolor}
\usepackage{stackengine}
\usepackage{multirow}
\usepackage{tabularx}
\usepackage{amsmath}
\usepackage{setspace}

\usepackage[%
    colorlinks=true,
    citecolor = blue,
    linkcolor= blue
]{hyperref}

\makeatletter

\def\@fnsymbol#1{\ensuremath{\ifcase#1\or \dagger\or \ddagger\or
   \mathsection\or \mathparagraph\or \|\or **\or \dagger\dagger
   \or \ddagger\ddagger \else\@ctrerr\fi}}
  \makeatother
\renewcommand{\footnote}{\fnsymbol{footnote}}

\title{Self-organization toward criticality by synaptic plasticity} 

\author{Roxana Zeraati $^{1,2}$, Viola Priesemann $^{3,4}$, Anna Levina $^{5,2,6,}$\footnote{anna.levina@uni-tuebingen.de}
	\\ {\small $^1$Graduate Training Center of Neuroscience, University of T\"ubingen, Germany} 
	 \\ {\small $^2$Max Planck Institute for Biological Cybernetics, T\"ubingen, Germany}
	 \\ {\small $^3$ Max Planck Institute for Dynamics and Self-Organization, G\"ottingen, Germany}
	 \\ {\small$^4$ Department of Physics, University of G\"ottingen, G\"ottingen, Germany}
	 \\ {\small $^5$ Department of Computer Science, University of T\"ubingen, T\"ubingen, Germany} 
	 \\ {\small $^6$ Bernstein Center for Computational Neuroscience T\"ubingen, T\"ubingen, Germany}
\date{}
	 }

\begin{document}
\maketitle

\section*{Abstract}
Self-organized criticality has been proposed to be a universal mechanism for the emergence of scale-free dynamics in many complex systems, and possibly in the brain. While such scale-free patterns were identified experimentally in many different types of neural recordings, the biological principles behind their emergence remained unknown. Utilizing different network models and motivated by experimental observations, synaptic plasticity was proposed as a possible mechanism to self-organize brain dynamics towards a critical point. In this review, we discuss how various biologically plausible plasticity rules operating across multiple timescales are implemented in the models and how they alter the network's dynamical state through modification of number and strength of the connections between the neurons. Some of these rules help to stabilize criticality, some need additional mechanisms to prevent divergence from the critical state. We propose that rules that are capable of bringing the network to criticality can be classified by how long the near-critical dynamics persists after their disabling. 
Finally, we discuss the role of self-organization and criticality in computation. Overall, the concept of criticality helps to shed light on brain function and self-organization, yet the overall dynamics of living neural networks seem to harnesses not only criticality for computation, but also deviations thereof.


\section{Introduction}


More than thirty years ago, Per Bak,  Chao Tang, and Kurt Wiesenfeld \cite{Bak1988} discovered a strikingly simple way to generate scale-free relaxation dynamics and pattern statistic, that had been observed in systems as different as earthquakes~\cite{gutenberg_seismicity_1941,gutenberg_earthquake_1956}, snow avalanches~\cite{birkeland_power-laws_2002}, forest fires~\cite{malamud_forest_1998}, or river networks~\cite{scheidegger1967bull,takayasu_new_1992}. Thereafter, hopes were expressed that this self-organization mechanism for scale-free emergent phenomena would explain how any complex system in nature worked, and hence it did not take long until the hypothesis sparked that brains should be self-organized critical as well~\cite{Beggs2003}. 

The idea that potentially the most complex object we know, the human brain, self-organizes to a critical state was explored early on by theoretical studies~\cite{chen_self-organized_1995,corral_self-organized_1995,herz_earthquake_1995,eurich_finite-size_2002}, but it took more than 15 years until the first scale-free “neuronal avalanches” were  discovered~\cite{Beggs2003}. Since then, we have seen a continuous, and very active interaction between experiment and theory. The initial, simple and optimistic idea that the brain is self-organized critical similar to a sandpile has been refined and diversified.
Now we have a multitude of neuroscience-inspired models, some showing classical self-organized critical dynamics, but many employing a set of crucial parameters to switch between critical and non-critical states ~\cite{eurich_finite-size_2002,Haldeman2005Critical,zierenberg_homeostatic_2018,cramer_control_2020}.  – Likewise the views on neural activity have been extended: We now have the means to quantify the distance to criticality even from the very few neurons we can record in parallel~\cite{wilting_inferring_2018}. Overall, we have observed in experiments, how developing networks self-organize to a critical state~\cite{yada_development_2017,tetzlaff_self-organized_2010,levina_subsampling_2017}, how states may change from wakefulness to deep sleep~\cite{Priesemann2013,lo_common_2004,allegrini_self-organized_2015,bocaccio_avalanche-like_2019,lombardi_critical_2020}, under drugs~\cite{meisel_antiepileptic_2020} or in a disease like epilepsy~\cite{scheffer_early-warning_2009,meisel_failure_2012,arviv_deviations_2016,hagemann_no_2020}. These results show how criticality and the deviations thereof can be harnessed for computation, but can also reflect cases where self-organization fails. 

Parallel to the rapid accumulation of experimental data, models describing the complex brain dynamics were developed to draw a richer picture. It is worthwhile noting that the seminal sandpile model~\cite{Bak_1987} already bears a striking similarity with the brain: The distribution of heights at each site of the system beautifully corresponds to the membrane potential of neurons, and in both systems, small perturbations can lead to scale-free distributed avalanches. However, whereas in the sandpile the number of grains naturally obeys a conservation law, the number of spikes or the summed potential in a neural network does not.


This points to a significant difference between classical SOC models and the brain: While in the SOC model the conservation law fixes the interaction between sites, in neuroscience connections strengths are ever-changing. Incorporating biologically plausible interactions is one of the largest challenges, but also the greatest opportunity for building the neuronal equivalent of a SOC model. Synaptic plasticity rules governing changes in the connections strengths often couple the interactions to the activity on different timescales. Thus, they can serve as the perfect mechanism for the self-organization and tuning the network's activity to the desired regime. 

Here we systematically review biologically plausible models of avalanche-related criticality with plastic connections. We discuss the degree to which they can be considered SOC proper, quasi-critical, or hovering around a critical state. We examine how they can be tuned towards and away from the classical critical state, and in particular, what are the biological control mechanisms that determine self-organization.  Our main focus is on models that exhibit signatures of criticality captured by the avalanche size distribution.


\section{Modeling neural networks with  plastic synapses}


Let us briefly introduce the very basics of neural networks, modeling neural circuits and synaptic plasticity. Most of these knowledge can be found in larger details in neuroscience text-books~\cite{Kandel,Dayanb,Gerstner_b}. The human brain contains about 80 billion neurons. Each neuron is connected to thousands of other neurons.  The connections between the neurons are located on fine and long trees of ``cables''. Each neuron has one such tree to collect signals from other neurons (\emph{dendritic tree}), and a different tree to send out signals to another set of neurons (\emph{axonal tree}). Biophysically, the connections between two neurons are realized by \emph{synapses}. These synapses are special: Only if a synapse is present between a dendrite and an axon can one neuron activate the other (but not necessarily conversely). The strength or weight $w_{ij}$ of a synapse  determines how strongly  neuron $j$ contributes to activating neuron $i$. If the summed input to a neuron exceeds a certain threshold within a short time window, the receiving neuron gets activated and fires a \emph{spike} (a binary signal). If a synapse $w_{ij}$ allows neuron $j$ to send signals to neuron $i$, it does not mean that the reverse synapse, $w_{ji}$ is also present. 
Thus, unlike classical physics systems, interactions between units are not symmetric but determined by a sparse, non-symmetric weight matrix $W$. Moreover, interactions are not continuous but pulse-like (\emph{spike}), and they are time-delayed by a few milliseconds: It takes a few milliseconds for a spike to travel along an axon, cross the synapse, and reach the cell body of the receiving neuron. Most interestingly, the synaptic weights $w_{ji}$ change over time. This is termed \emph{synaptic plasticity} and is the core mechanism behind learning.

Before we turn to studying synaptic plasticity in a model, the complexity of a living brain has to be reduced into a simplified model. Typically, neural networks are modelled with a few hundred or thousand of neurons. These neurons are either spiking, or approximated by ``rate neurons'' which represent the joint activity of an ensemble of neurons. Such rate neurons also exist \emph{in vivo}, e.g., in small animals, releasing graded potentials instead of spikes.
Of all neurons in the human cortex, 80\% are often modelled as excitatory neurons; when active, excitatory neurons contribute to activating their \emph{post-synaptic} neurons (i.e., the neurons to whom they send their signal). The other 20\% of neurons are inhibitory, bringing their post-synaptic neurons further away from their firing threshold. Effectively, an inhibitory neuron is modelled as having negative outgoing synaptic weights $w_{ij}$, whereas excitatory neurons have positive outgoing weights. In many simplified models, only one excitatory population is considered, and inhibition is implicitly assumed to be contributing to activity propagation probability that is already included in the excitatory connections. 
The connectivity matrix $W$ between the neurons is typically sparse, since most of the possible synapses are not realized. In models, the connectivity and initial strength of synapses are often drawn from some random distribution. In some studies, however, the impact of specific choices for connectivity and topology is explicitly explored, as outlined in this review (section \ref{sec:netowrk_wiring}). Finally, the model neurons often receive some external activation or input in addition to the input generated from the network connections to keep the network going and avoid an absorbing (quiescent) state. 

Numerous types of plasticity mechanisms shape the activity propagation in neuronal systems. One type of plasticity acts at the synapses regulating their creation and deletion, and determining changes in their weights $w_{ij}$. Thereby, regulating postsynaptic potentials, which govern the ability of the sending neuron to contribute to the activation of the receiving neuron and thus to activity propagation in the network. The other types of plasticity mechanisms regulate the overall excitability of the neuron, for example, by changing the spiking (activation) threshold or by adaptation currents.  

\begin{figure}[t]
    \centering\includegraphics[width=0.99\textwidth]{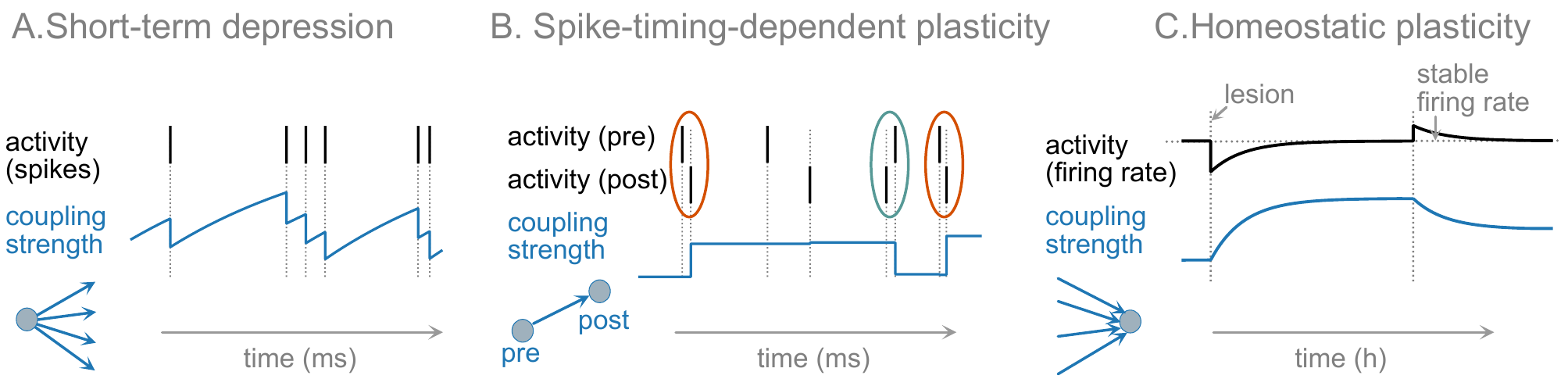}
    \caption{Schematic examples of synaptic plasticity. 
    \textbf{(A)} short-term synaptic depression acts on the timescale of spiking activity, and does not generate long-lasting changes. 
     \textbf{(B)} For spike-timing dependent plasticity (STDP), a synapse is potentiated upon causal pairing of pre- and postsynaptic activity (framed orange) and depressed upon anti-causal pairing (framed green), forming long-lasting changes after multiple repetitions of pairing.
     \textbf{(C)} Homeostatic plasticity adjusts presynaptic weights (or excitability) to maintain a stable firing rate. After reduction of a neuron's firing rate (e.g. after a lesion and reduction of input), the strengths of incoming excitatory synapses are increased to re-establish the neuron's target firing rate. In contrast, if the actual firing rate is higher than the target rate, then synapses are weakened, and the neuron returns to its  firing rate -- on the timescales of hours or days.}
    \label{fig:classification}
\end{figure}

The reasons and mechanisms of changing synaptic strength and neural excitability differ broadly. Changes of the synaptic strengths and excitability in the brain occur at different timescales that is particularly important for maintaining the critical dynamics. Some are very rapid acting within tens of milliseconds, or associated with every spike; others only make changes on the order of hours or even slower. For this review we simplified the classification in three temporally and functionally distinct classes, Figure~\ref{fig:classification}. 

The  timescale of a plasticity rule influences how it contributes to the state and collective dynamics of brain networks.
At the first level, we separate short-term plasticity acting on the timescale of dozens milliseconds, from the long-term plasticity acting with a time constant of minutes to days. As an illustration for short-term plasticity, we present prominent examples of short-term depression (see section~\ref{sec:short_term}). Among the long-term plasticity rules, we separate two distinct classes. 
First, plasticity rules that are explicitly associated with learning structures for specific activity propagation such as Hebbian and spike-timing-dependent plasticity (STDP, Figure~\ref{fig:classification}, middle). 
Second, homeostatic plasticity that maintains stable firing rate by up or down regulating neuronal excitability or synaptic strength to achieve a stable target firing rate over long time. This plasticity rule is particularly active after sudden or gradual changes in input to a neuron or neural network, and aims at re-establishing the neuron's firing rate (Figure~\ref{fig:classification}, right).   \\

\subsection{Criticality in network models}


Studying the distributions of avalanches is a common way to characterize critical dynamics in network models. Depending on the model, avalanches can be defined in different ways. When it is meaningful to impose the separation of timescales (STS), an avalanche is measured as the entire cascade of events following a small perturbation (e.g., activation of a single neuron) - until the activity dies out. However, a STS cannot be completely mapped to living neural systems due to the presence of spontaneous activity or external input. The external input impedes the pauses between avalanches and makes an unambigous separation difficult~\cite{Priesemann2014}. In models, such external input can be incorporated to make them more realistic.
To extract avalanches from living networks or from models with input, a pragmatic approach is chosen.
The binary events (spikes or thresholded continuous signals) of all channels or neurons are binned in time. An avalanche is then defined as a sequence of active bins between two silent bins.
In case the level of activity is so high that virtually no empty bins are observed, then the entire activity is thresholded, and only activity exceeding some threshold is considered as part of an avalanche.
While both binning and thresholding methods are widely used, concerns were raised that depending on the bin size \cite{Beggs2003,Priesemann2009,Priesemann2014,Priesemann2013}, the value of the threshold \cite{villegas2019timeseries}, or the intensity of input~\cite{das2018critical} distribution of observed avalanches and estimated power-law exponents might be altered. 
Therefore, to characterize critical dynamics using avalanches it is important to investigate the fundamental scaling relations between the exponents of avalanche size, duration and shapes to avoid misleading results \cite{Munoz_1999, friedman_universal_2012}, or instead use approaches to assess criticality that do not require the definition of avalanches~\cite{wilting_inferring_2018,Linkenkaer-Hansen_long-range_2001}. We elaborate on these challenges and bias-free solutions in a different book chapter~\cite{priesemann_assessing_2019}; for the remainder of this review, we assume that avalanches can be assessed unambiguously. 

\begin{figure}[t]
    \centering\includegraphics[width=0.85\textwidth]{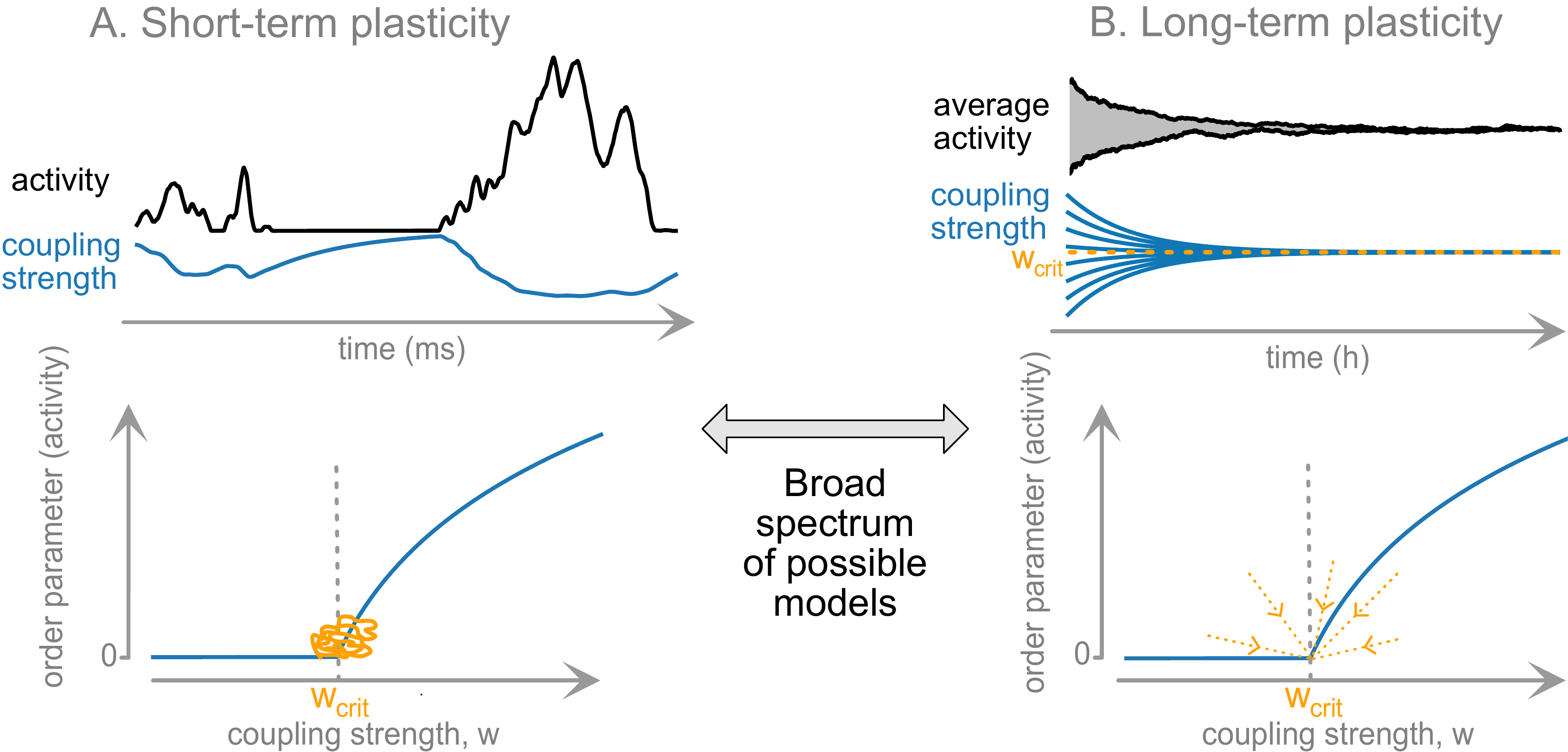}
    \caption{Classical plasticity rules and set-points of network activity.  \textbf{(A)} Short-term plasticity serves as immediate feedback (top). The resulting long-term behavior of the network hovers near the critical point (orange trace, bottom panel). \textbf{(B)} Long-term plasticity results in slow (timescale of hours or longer) convergence to the fixpoint of global coupling strength. In some settings, this fixpoint may correspond to the second-order phase-transition point (bottom), rendering the critical point a global attractor of dynamics.}    \label{fig:rationale}
\end{figure}

The timescale of a plasticity rule is crucial for the plasticity's ability in reaching and maintaining closeness to criticality. While short-term plasticity acts very quickly, it does not generate long-lasting, stable modifications of the network; and it can clearly serve as a feedback between activity and connection strength. Long-term plasticity, on the other side, takes longer to act, but can result in a stable convergence to critical dynamics, Figure~\ref{fig:rationale}.
To summarize their properties:
\begin{itemize}
\item Long-term plasticity is timescale-separated from activity propagation, whereas short-term plasticity evolves at similar timescales.
\item Long-term plasticity can 
self-organize a network to a critical state.
\item Short-term plasticity constitutes an inseparable part of the network dynamics.  It generates critical statistics in the data, working as a negative feedback.
\item The core difference: long-term plasticity, after convergence, can be switched off and the system will remain at criticality. Switching off short-term plasticity will almost surely destroy apparent critical dynamics.
\item There is a continuum of mechanisms on different timescales between these two extremes. Rules from this continuum can generate critical states that persist for varying time after rule-disabling, potentially even infinitely.\\
\end{itemize}

\subsection{Short-term synaptic plasticity} \label{sec:short_term}

The short-term plasticity (STP) captures activity-related changes in connection strength at a timescale close to the timescale of activity propagation, typically on the order of hundreds to thousands of milliseconds.  There are two dominant contributors to the short-term synaptic plasticity: the depletion of synaptic resources used for synaptic transmission and the transient accumulation of the {Ca$^{2+}$} ions that are entering the cell after each spike~\cite{Zucker2002}.

Synaptic resources are used for each spike, depleting a ready-releasable pool and leading to synaptic depression, i.e., decreasing the coupling strength after each spike, Figure~\ref{fig:classification}A. Synapses whose dynamics is dominated by depletion are called \emph{depressing synapses}~\cite{Markram1996}. At the same time, for some synapses, recent firing increases the probability of release for the vesicles in a ready-releasable pool. This mechanism leads to the increase of the coupling strength for a range of firing frequencies. Synapses with measurable contributions from it are called \emph{facilitating synapses}~\cite{Markram1998}. 

Although STP appears to be an inevitable consequence of synaptic physiology, multiple studies found that it can play an essential role in multiple brain functions. The most straightforward role is in the temporal filtering of  inputs, i.e., short-term depression will result in low-pass filtering~\cite{Abbott1997} that can be employed to reduce redundancy in the incoming signals~\cite{goldman2002redundancy}.  Additionally, it was shown to explain well the working memory~\cite{Mongillo2008}.

To model the changes in the connection strength associated with short-term synaptic plasticity it is sufficient to introduce two additional dynamic variables: $J_i$ indicates the number of synaptic resources available in neuron $i$, and $u_i$ fraction of these resources that will be used for spike. Coupling strength is captured by $w_i(t) = J_i(t) u_i(t)$ Each time when neuron $i$ emits
a spike at time $t_{\textrm{sp}}^{i}$, $J_i$ is reduced by $J_i(t_{\textrm{sp}}^{i}) u_i(t_{\textrm{sp}}^{i})$.  In between spikes, the resources recover and $J_{i}$ approaches
its resting value $J_{\mathrm{rest}}$ at a time scale $\tau_{J}$.
\begin{equation}
\dot{J}_{i}=\frac{1}{\tau_{J}}\left(J_{\mathrm{rest}}-J_{i}\right)-u_{i}J_{i}\delta\left(t-t_{\textrm{{sp}}}^{i}\right),\label{eq:jdynfas}
\end{equation}
with $\delta$ denoting Dirac delta function. 
To add synaptic facilitation, we equip $u_i$ with temporal dynamics, increasing it at each spike and decreasing between the spikes:
\begin{equation}
\dot{u}_{i}=\frac{1}{\tau_{u}}\left(u_{\mathrm{rest}}-u_{i}\right)+(1-u_{i}) u_{\mathrm{rest}}\delta\left(t-t_{\textrm{{sp}}}^{j}\right)\label{eq:udynfas}.\end{equation}

Including depressing synapses (Eq.~\ref{eq:jdynfas}) in the integrate-and-fire neuronal network was shown to increase the range of coupling parameters leading to the power-law scaling of avalanche size distribution~\cite{Levina_2007} as compared to the network without synaptic dynamics. If facilitation (Eq.~\ref{eq:udynfas}) is included in the model, an additional first-order transition arises~\cite{Levina2009}. Both models have an analytical mean-field solution. In the limit of the infinite network size, the critical dynamics is obtained for any large enough coupling parameter.  It was later suggested that the state reached by the system equipped with depressing synapses is not SOC, but self-organized quasi-criticality~\cite{bonachela2010self}, as it is not locally energy preserving. 

\begin{figure}[t]
    \centering\includegraphics[width=0.97\textwidth]{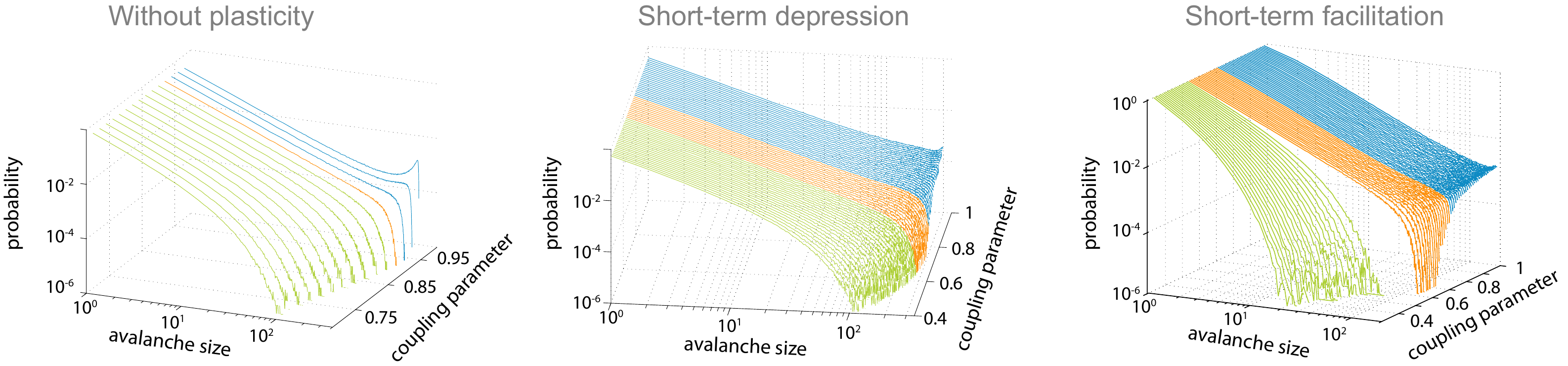}
    \caption{Short-term plasticity increases the range of the near critical regime. Left: model without plasticity reaches critical point only for single coupling parameter. Middle: short-term depression extends the range of parameters resulting in the critical dynamics. Right: short-term facilitation and depression together generate discontinuous transition.  }
    \label{fig:dyn_syn}
\end{figure}

The mechanism of the near-critical region extension with depressing synapses is rather intuitive. If there is a large event propagating through the network, the massive usage of synaptic resources effectively decouples the network. This in turn prevents the next large event for a while, until the resources are recovered. At the same time, series of small events allow to build up connection strength increasing the probability of large avalanche. Thus, for the coupling parameters above the critical values, the negative feedback generated by the synaptic depression allows to bring the system closer to the critical state. Complimentary, short-term facilitation can help to shift slightly subciritical systems to a critical state. 

The system with STD is essentially a two-dimensional dynamical system (with one variable corresponding to activity, and other to momentary coupling strength). Critical behavior is observed in the activity-dimension, over a long period of time while the coupling is hovering around the mean value as response to the changing activity. 
If the plasticity is ``switched off'' only a single parameter is generating critical-like dynamics. For the large system size, where even the smallest parameter deviation results in the big difference in the distribution, the probability to switch off plasticity at the moment of critical coupling strength is 0. Short-term plasticity generates correlations between consequent avalanche sizes and inter-avalanche intervals, similar to what was observed in the neuronal data~\cite{Lombardi2016}. 

After the first publication~\cite{Levina2006}, short-term depression was employed in multiple models discussing other mechanisms or different model for individual neurons. To name just few: in binary probabilistic networks~\cite{kinouchi2019stochastic}, in networks with long-term plasticity~\cite{michiels_van_kessenich_critical_2018,zeng_short-term_2019}, in spatially pre-structured networks~\cite{Wang2012}.  
In one of the few studies using leaky integrate-and-fire neurons, short term depression was also found to result in critical dynamics under a specific way of defining neuronal avalanches~\cite{Millman2010}. Later, it was shown that this particular definition will lead to power-law statistics also in clearly non-critical systems~\cite{,martinello2017neutral}.
In all cases the short-term plasticity contributes to the generation of a stable critical regime for a broad parameter range.   \\

\subsection{Long-term synaptic plasticity and network reorganization}

Long-term modifications in neuronal networks are created by two mechanisms: long-term synaptic plasticity and network reorganization. With the long-term plasticity, synaptic weights change over long timescales, but the adjacency matrix of the network remains unchanged. However, with network reorganization, new synapses are created, and some synapses might be removed. Both of these mechanisms can potentially contribute to self-organizing the network dynamics towards or away from criticality. 

Three types of long-term plasticity rules have been proposed as possible mechanisms for SOC: Hebbian plasticity, Spike-timing-dependent plasticity (STDP) and homeostatic plasticity. In Hebbian plasticity connections between near-synchronous neurons  are strengthened, while in STDP a temporally asymmetric rule is applied where depending on the order of pre- and post-synaptic spike-timings, connections can be strengthened or weakened. On the other hand, homeostatic plasticity acts as a negative feedback  that stabilizes the network's firing rate. In the following, we will discuss how each of these mechanisms can contribute to creating self-organized critical dynamics.\\


\subsubsection{Hebbian-like plasticity}

Hebbian plasticity is typically formulated in a slogan-like form: 
Neurons that fire together, wire together. This means that connections between neurons with similar spike-timing will be strengthened. This rule can imprint stable attractors into the network's dynamics, constituting the best candidate mechanism for memory formation. 
Hebbian plasticity in its standard form does not reduce coupling strength, thus without additional stabilization mechanisms Hebbian plasticity leads to runaway excitation. Additionally, presence of stable attractors makes it hard to maintain the scale-free distribution of avalanche sizes. 

The first papers uniting Hebbian-like plasticity and criticality came from Lucilla de Arcangelis' and Hans J. Herrmann's labs. In a series of publications, they demonstrated that a network of non-leaky integrators, equipped with plasticity and stabilizing synaptic scaling develops both power-law scaling of avalanches (with exponent 1.2 or 1.5 depending on the external drive) and power-law scaling of spectral density~\cite{dearcangelis_self-organized_2006, Pellegrini_2007}.
In the follow up paper, they realized multiple logical gates using additional supervised learning paradigm~\cite{arcangelis_learning_2010}. 

Using Hebbian-like plasticity to imprint patterns in the network and simultaneously maintain critical dynamics is a very non-trivial task. Uhlig et al.~\cite{Uhlig2013} achieved it by alternating Hebbian learning epochs with the epochs of normalizing synaptic strength to return to a critical state. The memory capacity of the trained network was close to the maximal possible capacity and remain close to criticality. However,  the network without homeostatic regulation towards a critical state achieved better retrieval. This might point to the possibility that classical criticality is not an optimal substrate for storing simple memories as attractors. However, in the so-far unstudied setting of storing memories as dynamic attractors, the critical system's sensitivity might make it the best solution. \\

\subsubsection{Spike-timing-dependent plasticity (STDP)}

Spike-timing-dependent plasticity (STDP) is a form of activity-dependent plasticity in which synaptic strength is adjusted as a function of timing of spikes in pre- and post-synaptic neurons. It can appear both in the form of long-term potentiation (LTP) or long-term depression (LTD)~\cite{sjostrom_spike-timing_2010}. Suppose the post-synaptic neuron fires shortly after the pre-synaptic neuron. In that case, the connection from pre- to the post-synaptic neuron is strengthened (LTP), but if the post-synaptic neuron fires after the pre-synaptic neuron, the connection is weakened (LTP), Figure~\ref{fig:classification} B. Millisecond temporal resolution measurements of pre- and postsynaptic spikes experimentally by Markram et al.~\cite{bi_synaptic_1998,markram_dendritic_1995,markram_regulation_1997} together with theoretical model proposed by Gerstner et al.~\cite{gerstner_neuronal_1996} put forward STDP as a mechanism for sequence learning. Shortly after that other theoretical studies~\cite{kempter_hebbian_1999,roberts_computational_2000, guyonneau_neurons_2005,farries_reinforcement_2007, costa_unified_2015} incorporated STDP in their models as a local learning rule.



Different functional forms of STDP are observed in different brain areas and across various species (for a review see~\cite{sjostrom_dendritic_2008}). For example, STDP in hippocampal excitatory synapses appear to have equal temporal windows for LTD and LTP~\cite{bi_synaptic_1998,nishiyama_calcium_2000,Zhang_1989}, while in neocortical synapses it exhibits longer LTD temporal windows~\cite{feldman_timing-based_2000, sjostrom_rate_2001}. Interestingly, an even broader variety of different STDP kernels were observed for  inhibitory connections~\cite{hennequin2017inhibitory}. 




The classical STDP is often modeled by modifying the synaptic weight $w_{ij}$ from pre-synptic neuron $j$ to post-synaptic neuron $i$ as
\begin{equation}
    \Delta w_{ij} = 
         \begin{cases}
      A_+(w_{ij})\, \text{exp}(\frac{t_j - t_i}{\tau_+}) & t_j<t_i\\
     - A_-(w_{ij})\, \text{exp}(\frac{t_j - t_i}{\tau_-}) & t_j\geq t_i
    \end{cases}  
    \label{equ:stdp_classic}
\end{equation}
where  $t_i$ and $t_j$ are latest spikes of neurons $i$ and $j$ and $\tau_+$ and $\tau_-$ are LTP and LTD time constants. Weight dependence functions $ A_+(w_{ij})$ and $A_-(w_{ij})$ control the synaptic weights to stay between $0$ and $w_{max}$, which is required from the biological point of view. Two families of weight dependence functions have been introduced: (i) soft weight bounds (multiplicative weights)~\cite{rossum_stable_2000}, (ii) hard weight bounds (additive weights)~\cite{gerstner_neuronal_1996}. Soft weight bounds are implemented as
\begin{equation}
    A_+(w_{ij}) = (w_{max} - w_{ij})\eta_+, \quad A_-(w_{ij}) = w_{ij}\eta_-\,,
    \label{equ:stdp_soft}
\end{equation}
where $\eta_+ <1$ and $\eta_- < 1 $ are positive constants. Weight dependence functions with hard bounds are defined using a Heaviside step function $H(x)$ as
\begin{equation}
    A_+(w_{ij}) = H(w_{max} - w_{ij})\eta_+, \quad A_-(w_{ij}) = H(-w_{ij})\eta_-\,.
    \label{equ:stdp_hard}
\end{equation}


There are two types of critical points that can be attained by networks with STDP.  
The first transition type is characterized by statistics of weights in the converged network. For instance, at this point synaptic coupling strengths~\cite{meisel_adaptive_2009} or the fluctuations in coupling strengths~\cite{shin_self-organized_2006} follow a power-law distribution. The second transition type is related to network's dynamics, it is characterized by presence of scale-free avalanches~\cite{rubinov_neurobiologically_2011,khoshkhou_spike-timing-dependent_2019, hernandez-urbina_self-organized_2017}. In these models STDP is usually accompanied by fine-tuning of some parameters or properties of the network to create critical dynamics. This suggests that STDP alone might not be sufficient for SOC.


\begin{figure}[t]
    \centering
     \includegraphics[trim=0 0 0 0, clip, width = 1\linewidth]{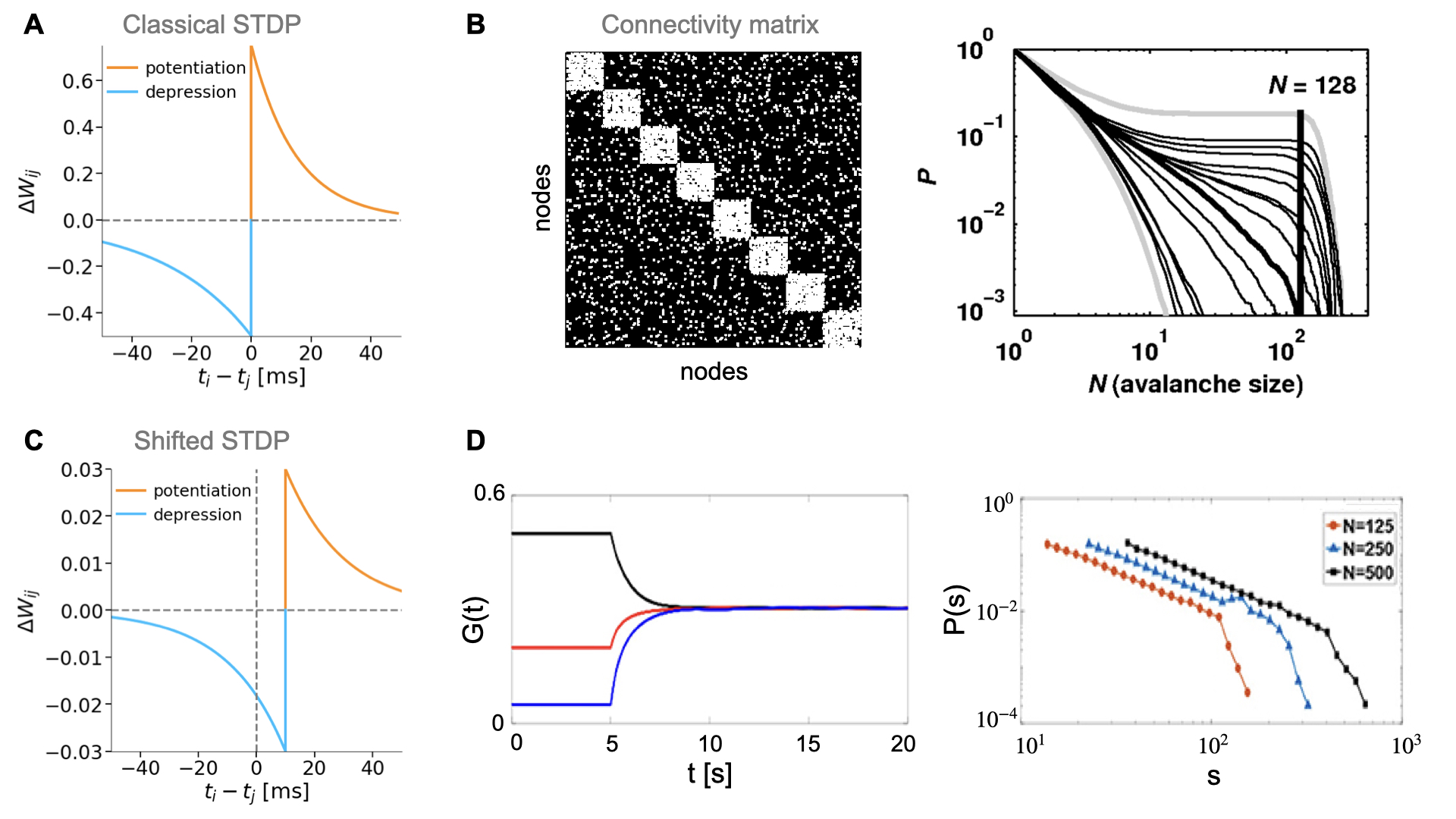}
    \caption{\textbf{Different STDP rules and their role in creating SOC.} \textbf{(A)} Classical STDP rule with asymmetric temporal windows. \textbf{(B)} A modular network that is rewired with a specific probability to create particular inter- and intra-modules connections (left) combined with classical STDP gives rise to dynamics characterized by power-law avalanche-size distribution (right, thick line). Non-power-law avalanche-size distributions correspond to other rewiring probabilities with the gray lines showing the two extremes of ordered and random networks (reproduced from~\cite{rubinov_neurobiologically_2011} under CC BY license). \textbf{(C)} Shifted STDP rule. \textbf{(D)} (left) Average coupling strength G(t) in a network with shifted STDP will converge to a steady state value. (right) Setting the STDP time-shit to $\tau = 10~$ms (equal to axonal delay time-constant) leads to emergence of power-law avalanche-size distributions that scale with the system size (reproduced from~\cite{khoshkhou_spike-timing-dependent_2019} under CC BY license).}
    \label{fig:stdp}
\end{figure}

Rubinov et al.~\cite{rubinov_neurobiologically_2011} developed a leaky integrate-and-fire (LIF) network model with modular connectivity (Figure~\ref{fig:stdp}A,B). In their model, STDP only gives rise to power-law distributions of avalanches when the ratio of connection between and within modules is tuned to a particular value.  Their results were unchanged for STDP rules with both soft and hard bounds.  However, they reported that switching off the STDP dynamics leads to the deterioration of the critical state, which disappears completely after a while. This property places the model in-between truly long-term and short-term mechanisms. Additionally, avalanches were defined based on the activity of modules (simultaneous activation of a large number of neurons within a module). In this modular definition of activity, SOC is achieved by potentiating within-module synaptic weights during module activation and depression of weights in-between module activations. While the module-based definition of avalanches could be relevant to the dynamics of cell-assemblies in the brain or more coarse-grained activity such as local field potentials (LFP), further investigation of avalanches statistics based on individual neurons activity is required. 


Observation of power-law avalanche distributions was later extended to a network of Izhikevich neurons with a temporally shifted soft-bound STDP rule~\cite{khoshkhou_spike-timing-dependent_2019} (Figure~\ref{fig:stdp}C,D). The shift in the boundary between potentiation and depression reduces the immediate synchronization between pre- and post-synaptic neurons that eventually stabilizes the synaptic weights and the post-synaptic firing rate similar to a homeostasis regulation~\cite{babadi_intrinsic_2010}. In the model, the STDP time-shift is set to be equal to the axonal delay time constant that also acts as a control parameter for the state of dynamics in the network. The authors showed that for a physiologically plausible time constant ($\tau = 10~$ms) network dynamics self-organizes to the edge of synchronization transition point. At this transition point, distribution of size and duration of avalanches follow a power-law-like distribution. They showed that the power-law exponents can be approximately fitted in the standard scaling relation required for a critical system~\cite{friedman_universal_2012}. However, since they defined avalanches based on thresholding of the global network firing, estimated avalanche distributions and fitted exponents might be generated by the thresholding~\cite{villegas2019timeseries}. \\

\subsubsection{Homeostatic regulations}

Homeostatic plasticity is a mechanism that regulates neural activity on a long timescale~\cite{Turrigiano_1998, lissin_activity_1998, obrien_activity-dependent_1998,turrigiano_homeostatic_2004,davis_homeostatic_2006,williams_homeostatic_2013}. 
In a nutshell, one assumes that every neuron has some intrinsic target activity rate.
Homeostatic plasticity then presents a negative feedback loop that maintains that target
rate and thereby stabilize network dynamics. In general, it
reduces (increases) excitatory synaptic strength or neural
excitability if the spike rate is above (below) a target rate, Figure~\ref{fig:classification}C. This mechanism
can stabilize a potentially unconstrained
positive feedback loop through Hebbian-type plasticity~\cite{bienenstock_theory_1982,miller_role_1994,abbott_synaptic_2000,turrigiano_hebb_2000,tetzlaff_synaptic_2011, zenke_synaptic_2013,keck_integrating_2017,zenke_hebbian_2017}. 
The physiological mechanisms of homeostatic plasticity are not fully disentangled yet. It can be implemented by a
number of physiological candidate mechanisms, such as
redistribution of synaptic efficacy~\cite{Markram1996,Tsodyks_1997}, synaptic scaling~\cite{Turrigiano_1998,lissin_activity_1998,obrien_activity-dependent_1998,fong_upward_2015}, adaptation of membrane excitability~\cite{davis_homeostatic_2006,pozo_unraveling_2010}, or
through interactions with glial cells~\cite{de_pitta_astrocytes_2016,virkar_feedback_2016}. 
Recent results highlight the involvement of
homeostatic plasticity in generating robust yet complex dynamics 
in recurrent networks~\cite{naude_effects_2013,gjorgjieva_homeostatic_2016,hellyer_local_2016}. 



\begin{figure}[t]
    \centering
     \includegraphics[trim=0 0 0 0, clip, width = 1\linewidth]{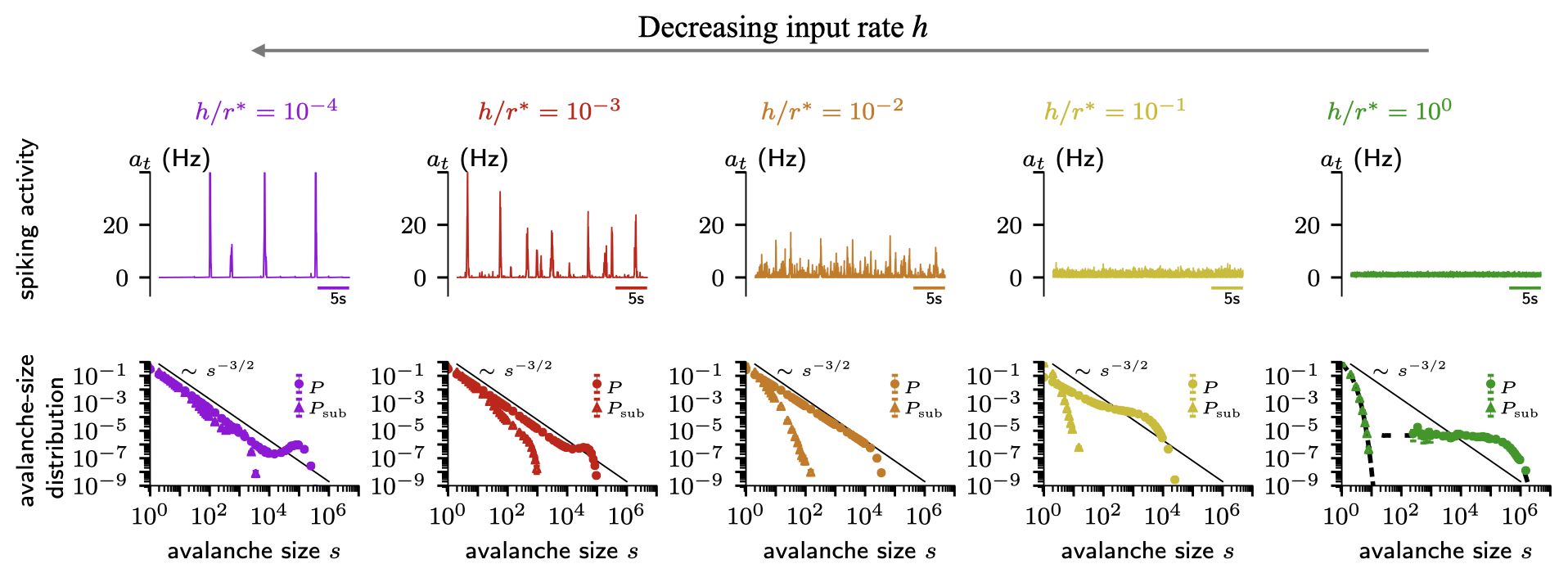}
    \caption{\textbf{Homeostatic plasticity regulation can create different types of dynamics in the network depending on input strength $h$, target firing rate $r^*$ and recurrent interactions.} (top) example spiking activity traces. (bottom) full-sampled (circles) and subsampled (triangles) avalanche size distributions averaged over 12 independent simulations. From left to right: network generates bursting ($m>1$, $h/r^* \leq 10^{-3}$ , purple-red), fluctuating ($m\approx 0.99$, $h/r^* \approx 10^{-2}$ and $m\approx 0.9$, $h/r^* \approx 10^{-1}$, orange-yellow) and irregular ($m\approx 0$, $h/r^* = 1$, green). Solid lines shows the corresponding branching process avalanche-size distributions $P(s) \propto s^{-3/2}$ \cite{harris_theory_1963} and dashed line is the analytical avalanche-size distribution of a Poisson process \cite{priesemann_can_2018} (reproduced from \cite{zierenberg_homeostatic_2018} by permission).}
    \label{fig:homes_model}
\end{figure}

In models, homeostatic plasticity was identified as one of the primary candidates to tune networks to criticality. The mechanism of it is straightforward: taking the analogy of the  branching process, where one neuron (or unit) on average activates $m$ neurons in the subsequent time step, the stable sustained activity that is the goal function of the homeostatic regulation requires $ m=m_c=1$ which is precisely the critical value~\cite{harris_theory_1963}. In 2007, Levina and colleagues made use of this principle. They devised a homeostasis-like rule, where all outgoing weights were normalized such that each neuron in the fully connected network activated on average $m=1$ neurons in the next time step~\cite{Levina_2007_CNS}.Thereby, the network tuned itself to a critical state.

Similar ideas have been proposed and implemented first in simple models and later also in
more detailed models. In the latter, homeostatic regulation tunes the ratio between excitatory and inhibitory synaptic strength ~\cite{poil_critical-state_2012,hellyer_local_2016,girardi-schappo_synaptic_2020,brochini_phase_2016,costa_self-organized_2017}.  
It then turned out  that due to the diverging temporal correlations, which emerge at criticality, the time-scale of homeostasis would also have to diverge~\cite{rocha_homeostatic_2018,brochini_phase_2016}. If the time-scale of the homeostasis is faster than the timescale of the dynamics, then the network does not converge to a critical point, but hovers around it, potentially resembling supercritical dynamics~\cite{rocha_homeostatic_2018,brochini_phase_2016,zierenberg_homeostatic_2018}. It is now clear that a self-organization to a critical state (instead of hovering around a critical state) requires that the timescale of homeostasis is slower than that of the network dynamics~\cite{rocha_homeostatic_2018,brochini_phase_2016,zierenberg_homeostatic_2018}.

Whether a system self-organizes to a critical state, or to a sub- or supercritical one is determined by a further parameter, which has been overlooked for a while: The rate of external input. This rate should be close to zero in critical systems to foster a separation of time scales~\cite{Priesemann2014,dickman_paths_2000}. 
Hence, basically all models that studied criticality were implemented with a vanishing external input rate.
In neural systems, however, sensory input and other brain areas provide continuous drive, and hence a separation of timescales is typically not realized~\cite{Priesemann2014}. As a consequence, avalanches merge, coalesce, and separate~\cite{Priesemann2014,zierenberg2020description,das2019critical}. It turns out that under homeostatic plasticity, the external input strength can become a control parameter for the dynamics~\cite{zierenberg_homeostatic_2018}: If the input strength is high, the system self-organizes to a subcritical state (Figure~\ref{fig:homes_model}, right). With weaker input, the network approaches a critical state (Figure~\ref{fig:homes_model}, middle). However when the input is too weak, pauses between bursts get so long that the timescale of the homeostasis again plays a role - and the network does not converge to a single state but hovers between sub-  and supercritical dynamics (Figure~\ref{fig:homes_model}, left). This study shows that under homeostasis the external input strength determines the collective dynamics of the network. 

Assuming that \emph{in vivo}, cortical activity is subject to some level of non-zero input, one expects a sightly subcritical state - which is indeed found consistently across different animals~\cite{zierenberg_homeostatic_2018, wilting_inferring_2018,hagemann_no_2020,ma_cortical_2019,wilting_between_2019}. In vitro systems, however, which lack external input, are expected to show bursty avalanche dynamics, potentially hovering around a critical point with excursions to supercriticality~\cite{zierenberg_homeostatic_2018,costa_self-organized_2017}. Such burst behavior is indeed characteristic for in vitro systems\cite{Beggs2003,tetzlaff_self-organized_2010,zierenberg_homeostatic_2018,friedman_universal_2012}.

\begin{figure}[t]
    \centering
     \includegraphics[trim=0 0 0 0, clip, width = 0.85\linewidth]{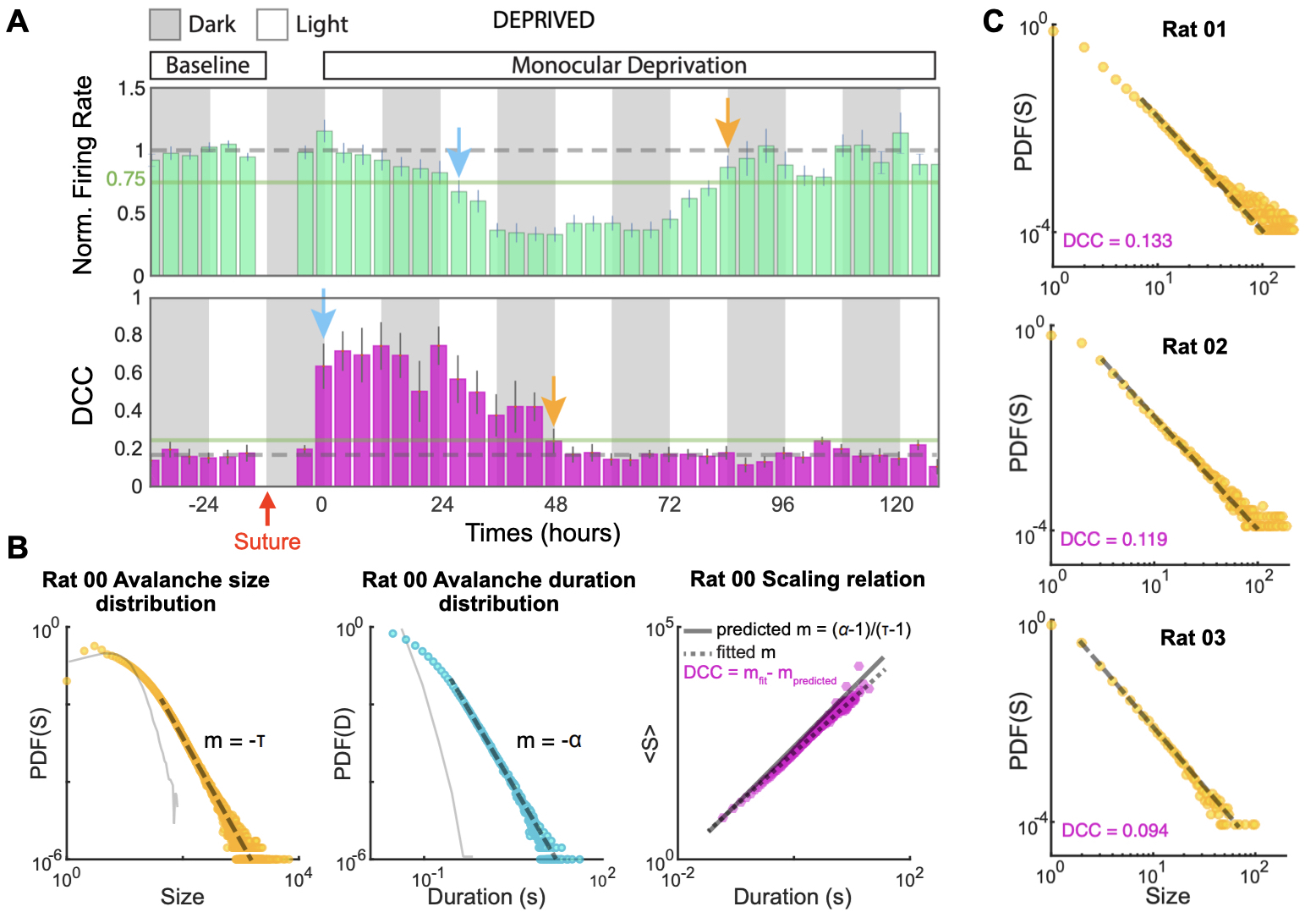}
    \caption{\textbf{Homeostatic regulations in visual cortex of rats tune the network dynamics to near criticality.} \textbf{(A)} (top) Firing rate of excitatory neurons during $7$ days of recording exhibit a biphasic response to monocular deprivation (MD). After $37$~h following MD firing rates were maximally suppressed (blue arrow) but came back to baseline by $84~$h (orange arrow). Rates are normalized to $24~$h of baseline recordings before MD. (bottom) Measuring the distance to criticality coefficient (DCC) in the same recordings. The mean DCC was significantly increased (blue arrow) upon MD, but was restored to baseline levels (near-critical regime) at $48~$h (orange arrow). \textbf{(B)} An example of estimation of DDC (right) using the power-law exponents from the avalanche-size distribution (left) and the avalanche-duration distribution (middle). Solid gray traces show avalanche distributions in shuffled data. DCC is defined as the difference between the empirical scaling (dashed gray line) and the theoretical value (solid gray line) predicted from the exponents for a critical system as the displayed formula \cite{friedman_universal_2012}. \textbf{(C)} Avalanche-size distributions and DCCs computed from $4~$h of single-unit data in three example animals show the diversity of experimental observations (reproduced from the the bioRxiv version of \cite{ma_cortical_2019} under CC-BY-NC-ND 4.0 international license).}
    \label{fig:homes}
\end{figure}



Recently, Ma and colleagues characterized in experiments how homeostatic scaling might re-establish close-to-critical dynamics \emph{in vivo} after perturbing sensory input~\cite{ma_cortical_2019} (Figure~\ref{fig:homes}). 
The past theoretical results would predict that after blocking sensory input in a living animal, the spike rate should diminish, and with the time-scale of homeostatic plasticity, a state close to critical or even super-critical would be obtained~\cite{zierenberg_homeostatic_2018,costa_self-organized_2017}. 
In a recent experiment, however, the behavior is more intricate. 
Soon after blocking visual input, the network became subcritical (branching ratio $m$ smaller than one~\cite{harris_theory_1963,wilting_inferring_2018}) and not supercritical. It then recovered to a close-to-critical state again within two days, potentially compensating the lack of input by coupling stronger to other brain areas.
The avalanche size distributions agree with the transient deviation to subcritical dynamics. This deviation to subcriticality is the opposite of what one might have expected under reduced input, and apparently cannot be attributed to concurrent rate changes (which otherwise can challenge the identification of avalanche distributions~\cite{priesemann_can_2018}):
The firing rate only started to decrease one day after blocking visual input.   
The authors attribute this delay in rate decay to excitation and inhibition reacting with different time constants to the blocking of visual input~\cite{ma_cortical_2019}.

Overall, although the exact implementation of the homeostatic plasticity on the pre- and postsynaptic sides of excitatory and inhibitory neurons  remains a topic of current research, the general mechanism allows for the long-term convergence of the system to the critical point, Figure~\ref{fig:classification}B. Homeostasis importantly contributes to many models including different learning mechanisms, stabilizing them.\\

\subsubsection{Network rewiring and growth  }
\label{sec:netowrk_wiring}

\begin{figure}[t]
    \centering
     \includegraphics[trim=0 0 0 0, clip, width = 1\linewidth]{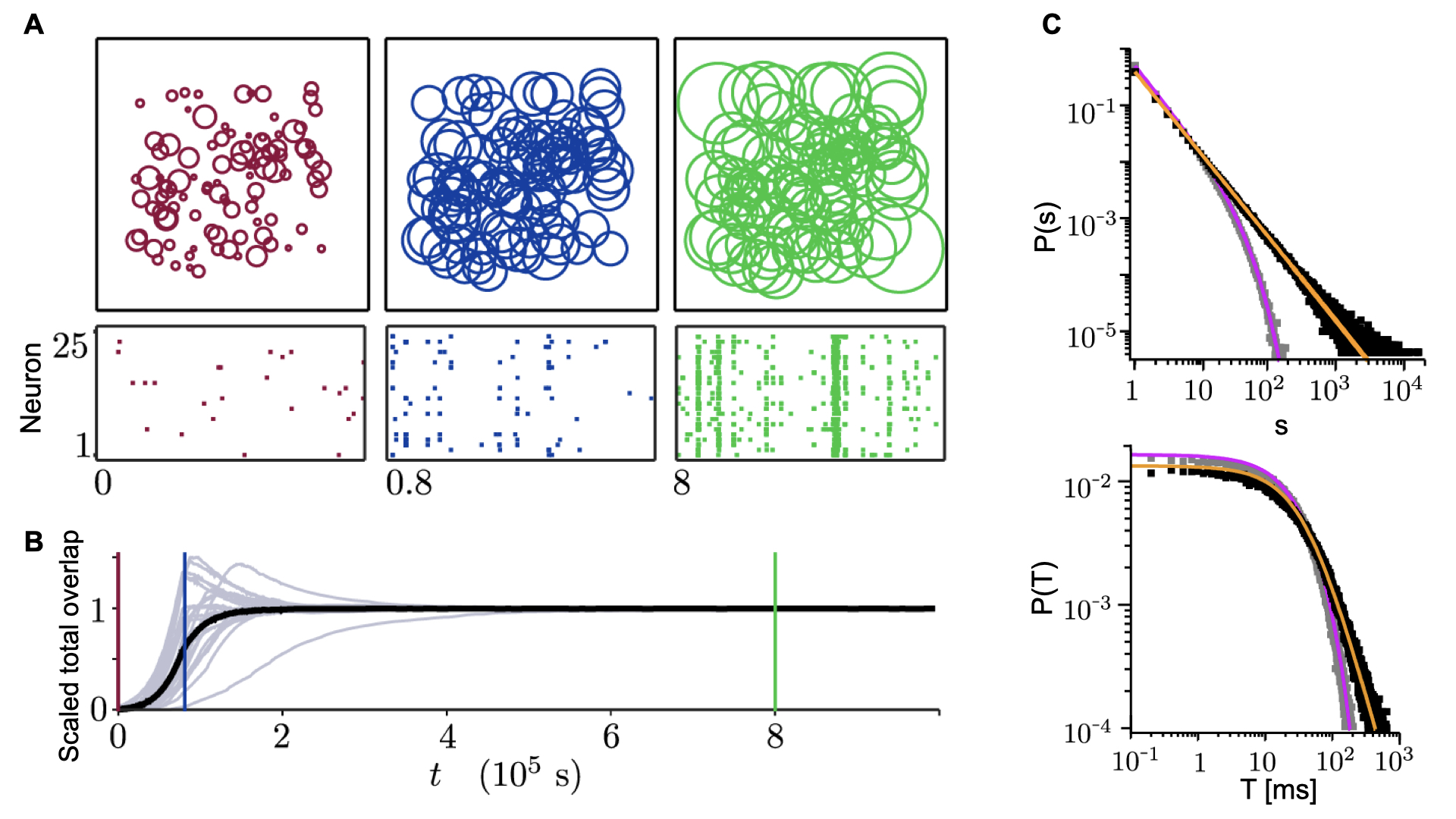}
    \caption{\textbf{Growing connections based on the homeostatic structural plasticity in a network model leads to SOC.} \textbf{(A)} Size of neurite fields (top) and spiking activity (bottom) change during the network growth process (from 25 sample neurons). From left to right: initial state (red), state with average growth (blue), stationary state reaching the homeostatic target rate (green). \textbf{(B)} Corresponding scaled total overlaps of 25 sample neurons (gray) and the population average (black) to the three different time points in (A). \textbf{(C)} Avalanche-size (top) and avalanche-duration (bottom) distributions. If the homeostatic target rate ($f_{target} = 2~$Hz) is significantly larger than the spontaneous rate ($f_{0} = 0.01~$Hz) both distributions follow a power-law (black: simulation,  orange: analytic). Small homeostatic target rate ($f_{target} = 0.04~$Hz) leads to subcritical dynamics (gray: simulation, pink: analytic) (reproduced from \cite{kalle_kossio_growing_2018} with permission).}
    \label{fig:growth}
\end{figure}

Specific network structures such as small-world~\cite{lin_self-organized_2005,Levina_2007,de_arcangelis_self-organized_2002,dearcangelis_self-organized_2006} or scale-free~\cite{fronczak_self-organized_2006,bianconi_clogging_2004,hughes_solar_2003,paczuski_heavenly_2004,michiels_van_kessenich_critical_2018,arcangelis_learning_2010} networks were found to be beneficial for the emergence of critical dynamics.  These network structures are in particular interesting since they have been also observed in both anatomical and functional brain networks~\cite{bassett_small-world_2006,bullmore_complex_2009,he_scale-free_2014,eguiluz_scale-free_2005}. To create such topologies in neural networks long-term plasticity mechanisms have been used. For instance, scale-free and small-world structures emerge as a consequence of STDP between the neurons~\cite{shin_self-organized_2006}. In addition, Hebbian plasticity can generate small-world networks~\cite{siri_effects_2007}.

Another prominent form of network structures are hierarchical modular networks (HMN) that can sustain critical regime for a broader range of control parameters~\cite{Moretti2013, Wang2012,rubinov_neurobiologically_2011}. Unlike a conventional critical point where control parameter at a single value leads to scale-free avalanches, in HMNs power-law scaling emerges for a wide range of parameters. This extended critical-like region can correspond to a Griffits phase in statistical mechanics~\cite{Moretti2013}. Different rewiring algorithms have been proposed to generate HMN from an initially  randomly connected~\cite{Wang2012} or a fully connected modular network~\cite{Moretti2013,rubinov_neurobiologically_2011}.

Experimental observations in developing neural cultures suggest that  connections between neurons grow in a way such that the dynamics of the network eventually self-organizes to a critical point (i.e., observation of scale-free avalanches)~\cite{yada_development_2017,tetzlaff_self-organized_2010}. Motivated by this observation, different models have been developed to explain how neural networks can grow connections to achieve and maintain such critical dynamics using homeostatic structural plasticity~\cite{van_ooyen_activity-dependent_1994,van_ooyen_complex_1996,Abbott_2007,tetzlaff_self-organized_2010,kalle_kossio_growing_2018,droste_2013_analytical, landmann_self-organized_2020} (for a review see~\cite{van_ooyen_homeostatic_2019}). In addition to homeostatic plasticity, other rewiring rules inspired by Hebbian learning were also proposed to bring the network dynamics towards criticality~\cite{bornholdt_topological_2000,bornholdt_self-organized_2003,rybarsch2014avalanches}. However, implementation of such network reorganizations seems to be less biologically plausible.

In most of the models with homeostatic structural plasticity, the growth of neuronal axons and dendrites is modeled as an expanding (or shrinking) circular neurite field. The growth of the neurite field for each neuron is defined based on the neuron's firing rate (or internal {Ca$^{2+}$} concentration). A firing rate below the homeostatic target rate ($f_{target}$) expands the neurite field, and a firing rate above the homeostatic target rate shrinks it. In addition, when neurite fields of two neurons overlap a connection between them will be created with a strength proportional to the overlapped area. Kossio et al.~\cite{kalle_kossio_growing_2018} showed analytically that if the homeostatic target rate is significantly larger than the spontaneous firing rate of the network, such growth mechanism would bring the network dynamics to a critical point with scale-free avalanches (Figure~\ref{fig:growth}). However, for a small target rate subcritical dynamics will arise. 

Tetzlaff et al.~\cite{tetzlaff_self-organized_2010} proposed a slightly different mechanism where two neurites fields are assigned separately for axonal growth and dendritic growth to each neuron. While changes in the size of dendritic neurite fields follows the same rule as explained above, neurite fields of axons follow an exact opposite rule. The model simulations start with all excitatoryry neurons, but in the middle phase $20\%$ of the neurons are changed into inhibitory ones. This switch is motivated by the transformation of GABA neurotransmitters from excitatory to inhibitory during development~\cite{ben-ari_gaba_2012}. They showed that when the network dynamics converge to a steady-state regime, avalanche-size distributions follow a power-law.

\section{Hybrid Mechanisms of Learning and Task Performance}
In living neural networks, multiple plasticity mechanisms occur simultaneously. The joint contribution of diverse mechanisms has been studied in the context of criticality in a set of models~\cite{michiels_van_kessenich_critical_2018,peng2013attaining,stepp_synaptic_2015}. 
A combination with homeostatic-like regulation is typically necessary to stabilize Hebbian or spike-timing-dependent plasticity (STDP), e.g., learning binary tasks such as an XOR rule with Hebbian plasticity~\cite{michiels_van_kessenich_critical_2018} or sequence learning with STDP~\cite{scarpetta_effects_2013,scarpetta2014alternation,scarpetta_hysteresis_2018,loidolt_sequence_2020,papa_criticality_2017}. These classic plasticity rules have been paired with regulatory normalization of synaptic weights to avoid a self-amplified destabilization~\cite{zenke_hebbian_2017,zenke_synaptic_2013,keck_integrating_2017}. Additionally, short-term synaptic depression stabilizes the critical regime, and if it is augmented with meta-plasticity~\cite{peng2013attaining} the stability interval is increased even further, possibly allowing for stable learning.



In a series of studies, Scarpetta and colleagues investigated how sequences can be memorized by STDP, while criticality is maintained~\cite{scarpetta_effects_2013,scarpetta_hysteresis_2018,scarpetta2014alternation}. By controlling the excitability of the neurons, they achieved a balance between partial replays and noise resulting in power-law distributed avalanche sizes and durations~\cite{scarpetta_effects_2013}. They later reformulated the model and used the average connection strength as a control parameter, obtaining similar results~\cite{scarpetta_hysteresis_2018,scarpetta2014alternation}. Whereas STDP fosters the formation of sequence memory, Hebbian plasticity is known to form assemblies (associations), and in the Hopfield network enables memory completion and recall~\cite{Hopfield_1982}. 
A number of studies showed that the formation of such Hebbian ensembles is also possible while maintaining critical dynamics~\cite{scarpetta_neural_2013,scarpetta_hysteresis_2018,Uhlig2013}. These studies show that critical dynamics can be maintained in networks, which are learning classical tasks.


The critical network can support not only memory but also real computations such as performing logical operations (OR, AND or even XOR) \cite{michiels_van_kessenich_critical_2018, arcangelis_learning_2010}. To achieve this, the authors build upon the model with Hebbian-like plasticity that previously shown to bring the network to a critical point~\cite{dearcangelis_self-organized_2006}.  They added the central learning signal~\cite{bak_adaptive_2001}, resembling dopaminergic neuromodulation. Authors demonstrated both with~\cite{michiels_van_kessenich_critical_2018} and without~\cite{arcangelis_learning_2010} short-term plasticity that the network can be trained to solve XOR-gate task.  

These examples lead to the natural question of whether criticality is always optimal for learning. The criticality hypothesis attracted much attention, precisely because models at criticality show properties supporting optimal task performance. A core properties of criticality is a maximization of the dynamic range~\cite{Kinouchi_optimal_2006,zierenberg2020tailored}, the sensitivity to input, and diverging spatial and temporal correlation lengths~\cite{stanley_introduction_1972,sethna_statistical_2006}.
In recurrent network models and experiments, such boost of input sensitivity and memory have been demonstrated by tuning networks systematically towards and away from criticality~\cite{Shew2009,Shew2012,shew2013,boedecker2012,boedecker_modeling_2013,Kinouchi_2006,bertschinger2004}. 


When not explicitly incorporating a mechanism that drives the network to criticality, learning networks can be pushed away from criticality to a subcritical regime~\cite{lazar_sorn_2009,delpapa_fading_2019,papa_criticality_2017,cramer_control_2020}.
This is in line with the results above that networks with homeostatic mechanisms become subcritical under increasing network input (Figure~\ref{fig:homes_model}).
Subcritical dynamics might indeed be favorable when reliable task performance is required, as the inherent variability of critical systems may corroborate performance variability~\cite{Priesemann2014,cocchi_criticality_2017,nolte_cortical_2019,gollo_coexistence_2017,wilting_operating_2018,wilting_25_2019,Tomen_Marginally_2014}.

Recently, the optimal working points of recurrent neural networks on a neuromorphic chip were demonstrated to depend on task complexity~\cite{cramer_control_2020}.
The neuromorphic chip implements  spiking integrate-and-fire neurons with STDP-like depressive plasticity and slow homeostatic recovery of synaptic strength. It was found that complex tasks, which require integration of information over long time-windows, indeed profit from critical dynamics, whereas for simple tasks the optimal working point of the recurrent network was in the subcritical regime~\cite{cramer_control_2020}. 
Criticality thus seems to be optimal particularly when a task makes use of this large variability, or explicitly requires the long-range correlation in time or space, e.g. for active memory storage.



\section{Discussion}

We summarized how different types of plasticity contribute to the convergence and maintenance of the critical state in neuronal models. The short-term plasticity rules were generally leading to hovering around the critical point, which extended the critical-like dynamics for an extensive range of parameters. The long-term homeostatic network growth and homeostatic plasticity, for some settings, could create a global attractor at the critical state. Long-term plasticity associated with learning sequences, patterns or tasks required additional mechanisms (i.e. homeostatic) to maintain criticality. 


The first problem with finding the best recipe for criticality in the brain is our inability to identify the brain's state from the observations we can make. We are slowly learning how to deal with strong subsampling (under-observation) of the brain network~\cite{wilting_inferring_2018, levina_subsampling_2017,Priesemann2009,ribeiro2010,spitzner_mr_2020,zeraati_estimation_2020}. However, even if we obtained a perfectly resolved observation of all activity in the brain, we would face the problem of constant input and spontaneous activation that renders it impossible to find natural pauses between avalanches, and hence makes avalanche-based analyses ambiguous~\cite{Priesemann2014}.
Hence, multiple avalanche-independent options were proposed as alternative assessments of criticality in the brain:
(i.) detrended fluctuation analysis~\cite{Linkenkaer-Hansen_long-range_2001} allows to capture the scale-free behavior in long-range temporal correlations of EEG/MEG data, 
(ii.) critical slowing down~\cite{meisel_critical_2015} suggests a closeness to a bifurcation point, 
(iii.) divergence of susceptibility in the maximal entropy model fitted to the neural data~\cite{Tkacik_thermodynamics_2015}, or the renormalization group approach~\cite{meshulam_coarse_2019} indicates a closeness to criticality in the sense of thermodynamic phase-transitions, and 
(vi.) estimating the branching parameter directly became feasible even from a small set of neurons; this estimate returns a quantification of the distance to criticality~\cite{wilting_inferring_2018}.
Finding the best way to unite these definitions, and select the most suitable ones for a given experiment remains largely an open problem. 



Investigating the criticality hypothesis for brain dynamics has strongly evolved in the past decades, but is far from being concluded. On the experimental side, sampling limits our access to collective neural dynamics~\cite{levina_subsampling_2017,neto_unified_2020}, 
and hence it is not perfectly clear yet how close to a critical point different brain areas operate. For cortex in awake animals, evidence points to a close-to-critical, but slightly subcritical state~\cite{wilting_between_2019,ma_cortical_2019,hagemann_no_2020}. 
The precise working point might well depend on the specific brain area, cognitive state and task requirement~\cite{wilting_operating_2018,Tomen_Marginally_2014,shriki2013neuronal,shew2013,shew_adaptation_2015,clawson_adaptation_2017,yu2017maintained,cramer_control_2020,carhart-harris_neural_2016,simola_critical_2017, tagliazucchi_breakdown_2013}. Thus instead of self-organizing precisely to criticality, the brain could make use of the divergence of processing capabilities around the critical point. Thereby, each brain area might optimize its computational properties by tuning itself towards and away from criticality in a flexible, adaptive manner~\cite{wilting_operating_2018}. In the past decades, the community has revealed the local plasticity rules that would enable such a tuning and adaption of the working point. Unlike classical physics systems, which are constrained by conservation laws, the brain and the propagation of neural activity is more flexible and hence can adhere in principle a large repertoire of working points - depending on task requirements. 


Criticality has been very inspiring to understand brain dynamics and function. We assume that being perfectly critical is not an optimal solution for many brain areas, during different task epochs. However, studying and modelling brain dynamics from a criticality point of view allows to make sense of the high-dimensional neural data, its large variability,  and to formulate meaningful hypothesis about dynamics and computation, many of which still wait to be tested.\\






\section*{Conflict of interest statement}
The authors declare that the research was conducted in the absence of any commercial or financial relationships that could be construed as a potential conflict of interest.

\section*{Author contributions}
RZ, VP, and AL designed the research. RZ and AL prepared the figures. All authors contributed to writing and reviewing the manuscript.

\section*{Acknowledgments}
We are very thankful to Sina Khajehabdollahi, Manos Giannakakis, and Sahel Azizpour for reading the initial version of the manuscript and their constructive comments.
\section*{Funding}
This work was supported by a Sofja Kovalevskaja Award
from the Alexander von Humboldt Foundation, endowed by the Federal Ministry of Education and Research (RZ, AL), Max Planck Society (VP), \textbf{SMART}\textit{START} 2 program provided by Bernstein Center for Computational Neuroscience and Volkswagen Foundation (RZ). We acknowledge the  support from the BMBF through the T\"ubingen AI Center (FKZ: 01IS18039B).


\end{document}